\documentclass[aps,pre,preprint,groupedaddress]{revtex4-2}
\usepackage{graphicx}
\usepackage{amsmath}
\usepackage{amssymb}
\usepackage[colorlinks,citecolor=blue,urlcolor=blue,linkcolor=blue]{hyperref}

\begin{document}


\title{Collisional regime during the discharge of a 2D silo}


\author{Roberto Ar\'evalo}
\email[]{rarevalo@fcirce.es}
\affiliation{Research Centre for Energy Resources and Consumption (CIRCE), Ave. Ranillas 3D, 1st floor, 50018 Zaragoza, Spain.}


\date{\today}

\begin{abstract}
The present work reports a novel investigation into the collisional dynamics of
particles in the vicinity of the outlet of a \textit{2D} silo using molecular
dynamics simulations. Most studies on this granular system focus in the bulk of
the medium. In this region contacts are permanent or long-lived, so continuous
approximations are able to yield results for velocity distributions or mass flow.
Close to the exit, however, the density of the medium decreases and contacts are
instantaneous. Thus, the collisional nature of the dynamics becomes significant,
warranting a dedicated investigation as carried out in this work. More interesting,
the vicinity of the outlet is the region where the arches that block the flow for
small apertures are formed. It is found that the transition from the clogging
regime (at small apertures) to the continuous flow regime is smooth in collisional
variables. Furthermore, the dynamics of particles as reflected by the distributions of
the velocities is as well unaffected. This result implies that there is no critical
outlet size that separates both regimes, as had been proposed in the
literature. Instead, the results achieved support the alternative picture in which
a clog is possible for any outlet size.
\end{abstract}


\maketitle

\section{Introduction}
\label{intro}

\begin{figure}
\begin{center}
\includegraphics[width=\textwidth, keepaspectratio=true]{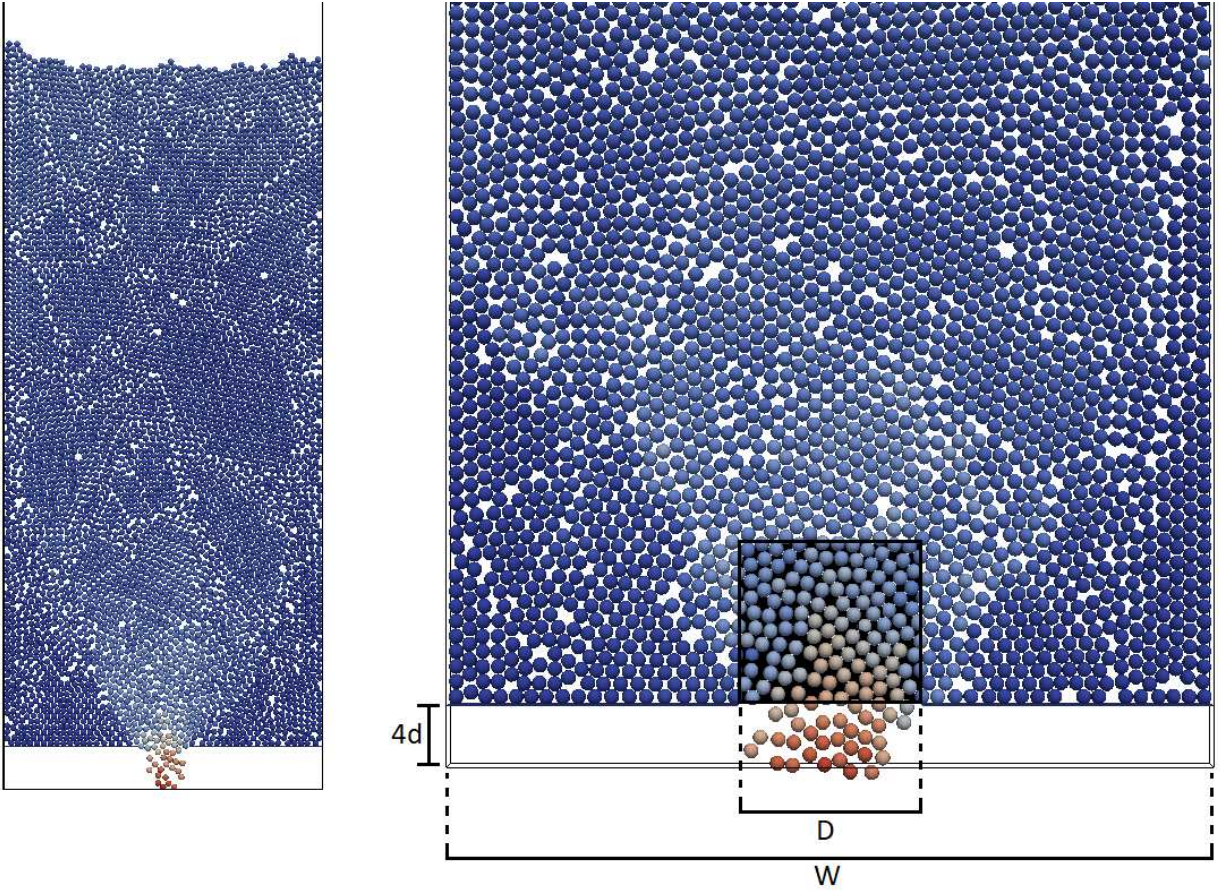}
\end{center}
\caption{{\bf Snapshot of a simulation.}
\textit{Left}: A silo with a large aperture size flowing in steady state.
\textit{Right}: Region close to the exit. The shadowed square represents the
$D\times D$ observation window.
} \label{fig1}
\end{figure}

The discharge of a silo by gravity has been during years a favourite process to
explore the dynamics of granular materials~\cite{To,Lix,Alonso1,Zhou,Yang,Alonso2,Ashour,Fullard,Wan,Cruz,Liu,Calderon,Peralta,Darias,Angel}.
These media consist of macroscopic particles interacting by frictional contact
forces, giving rise to a plethora of novel behaviours studied by engineers and
physicists alike~\cite{Edwards,Kadanoff,Vahidi,Levay,Dijksman,Baldovin,Grasselli,Gonzalez1,Li,Saleh,Tian}.
In a flat-bottomed silo (see Fig~\ref{fig1}) a column of particles or
grains is allowed to come to rest before opening a hole at the base. Then, a stream
of particles outflows the silo pulled by the force of gravity. Depending on the
relation between the size of the particles $d$ and that of the exit $D$, two main
regimes can be identified. When particles are several times smaller than the exit,
the flow is smooth and continuous. Like the one in an hourglass, it proceeds
unperturbed and reminds the outflow of a liquid. However, due to the Janssen
effect~\cite{Janssen}, the pressure is constant for most of the depth of the silo,
and the flow is independent of the height of material inside the silo~\cite{Beverloo}.

Upon reducing the size of the outlet ($D\lesssim 5d$) the dynamics of the flow starts to change.
In ~\cite{Unac} a sharp transition in the characteristic frequency of flow
fluctuations is found and related to the stability of transitory arches even well
inside the continuous regime. When the size of the particles becomes comparable to
that of the exit the flow starts to exhibit disturbances. In two dimensions, this
happens when the ratio $D/d$ of hole to grain size is around $5$. Oscillations
appear which grow upon further reducing the outlet size. Eventually, the flow is
arrested by the formation of an arch of particles. Around $D/d=5$ many arches are
short-lived, being soon dragged by the stream of incoming particles. This results
in an intermittent flow regime. But below that value stable arches appear that is
necessary to remove to resume the flow.

In the bulk of the silo particles present a relatively high packing density and most
contacts are permanent, rather than exhibiting a collisional regime. Under these
conditions, the flow can be modelled as a continuous medium~\cite{Nedderman1,Tuzun}
to predict the shape of the velocity profile with excellent accuracy. Approaching
the outlet, however, the situation changes. The packing fraction of the particles
decreases fast, and the discontinuous nature of the medium becomes more important.
The low density precludes the existence of durable contacts, so the dynamics
of particles is now dominated by collisions. Although a continuous description
in which the relevant length scale is $D$ is possible~\cite{Rubio}, high-speed
images show the particles colliding in a way that reminds one of a gas until
rather suddenly an arch is formed~\cite{Zuriguel_obs}.

These two flow regimes, continuous at large values of $D/d$ and clogging at low
values, are reflected in the behaviour of the mass flow of particles. In the
continuous regime, the mass flow is proportional to $D^{5/2}$ in $3D$ or $D^{3/2}$
in $2D$. This scaling can be simply rationalized as the product of the area of
the outlet times the typical velocity of the particles when they reach it. However,
in the clogging regime the scaling of the flow grows faster. This is due to the
variation of the packing fraction, which grows for small values of $D/d$ until it
reaches a constant value. From the point where the packing fraction saturates the
expected power law behaviour of the flow is observed~\cite{Mankoc}.

A point that has been debated is whether exists a critical $D/d$ ratio that
separates the clogging regime, in which the flow is arrested by arches, from the
continuous regime~\cite{Zuriguel_3d,Janda_epl,To2,Durian_prl}. The argument in
favour is that the number of grains out-poured from the silo between two clogs
(termed an avalanche) seems to grow as a critical power law of $D/d$. However,
these measurements are rather difficult to carry out due to the fast growth of the
avalanches of grains. As a consequence, the data can be fitted equally well to
functions of $D$ which present a critical value and to functions that do not. By
other side, the critical exponent and the value of the critical outlet size in $2D$ 
($D_c\sim 8.5d$~\cite{Janda_epl}) turn out to be rather large to be easily interpretable. 
The recent work~\cite{Durian_prl} introduces the concept of clogging configurations to build
a strong statistical argument. The conclusion is that the probability of forming an
arch expanding the outlet size simply decreases until it is unobservable within
experimental time windows.

The aim of the present work is to study the gas-like behaviour of particles in the
vicinity of the outlet during the transition from the clogging to the continuous
flow regime. Unlike the approaches mentioned, ours takes at face value the
collisional nature of the dynamics close to the outlet. It is found that the
transition is smooth in $D/d$ indicating that there is no critical outlet size
separating both regimes. This conclusion is reinforced studying the collisional
dynamics of the particles in the vicinity of the outlet.

\section{Materials and methods}
\subsection{Numerical method}
\label{method}

The present study carries out molecular dynamics simulations of granular materials
using the free and open source software \textit{LIGGGHTS}~\cite{liggghts} which is an
extension for granular matter of the the well-known \textit{LAMMPS}~\cite{lammps} package. The
advantage of this choice (apart of being open and free) is that \textit{LIGGGHTS} has many
interaction potentials already built-in, including the most common for granular
matter. Besides, it is efficiently implemented and parallelized for fast simulations.

The particles are modelled as monosized spheres of diameter $d=1mm$. In this way we
avoid introducing size segregation in the silo, which would complicate the analysis
and obscure the phenomena under study. The width $W$ of the silo is $50d$ to avoid
the influence of the position of the walls~\cite{Rapaport}. The height is $100d$
for small to medium outlet size, $200d$ for the larger outlets, and $220d$ for the 
case $D=15d$. This height ensures that the dynamics is independent of the filling. 
Finally, a depth of $1.1d$ was used as is the case of some experiments~\cite{Janda}. 
The number of particles used ranges from $5000$ to $10000$.

In the present work we consider only contact forces, under this condition two
particles interact whenever their distance $r_{ij}$ becomes smaller than the sum
of their radii. To model the interaction between particles we choose the Hertz
contact, which includes both frictional and dissipative terms:

\begin{eqnarray}
        \mathbf{F}_n &=& k_{n}\xi\mathbf{n}_{ij} -\gamma_{n}v_{i,j}\mathbf{n}_{ij}\label{normal}\\
        \mathbf{F}_t &=& k_{t}\xi\mathbf{t}_{ij} -\gamma_{t}v_{i,j}\mathbf{t}_{ij}\label{tangential}
\end{eqnarray}

The first equation is the component of the force in the normal direction of the
impact. The spring force is proportional to the overlap
$\xi=\frac{1}{2}(d_i+d_j)-r_{ij}$ of the particles. The second term is a damping
proportional to the relative velocity of the colliding particles. The parameter
$k_{n}$ is the elastic constant and $\gamma_{n}$ is a viscoelastic damping constant
whose role is to dissipate energy during the collision. Analogously, the force in
the tangential direction depends on a restoring term, proportional
to the sliding of the particles, and a damping term that dissipates energy. The
damping constants are related to the restitution coefficient (see definition below).
They control how much energy is lost in a collision and, thus, affect at how fast
the medium settles in a static state, \emph{e. g.}, when the silo is filled.
Particles with a large restitution coefficient (large damping) separate more after
a collision, however due to the high density of the medium the steric effects are
more important than the influence of a single parameter on the overall behavior
of the medium. The tangential overlap $\xi\mathbf{t}_{ij}$ increases while the
contact lasts but is truncated to fulfil the Coulomb criterium $F_t\leq \mu F_n$,
where $\mu$ is the friction coefficient.

For the simulation we fix the values of the Young's modulus $Y$, Poisson ratio $\nu$,
coefficient of restitution $e$ and friction coefficient $\mu$. These are related
to the constants of the force model as follows:

\begin{eqnarray}
        k_n &=& \frac{4}{3} Y^*\sqrt{R^*\xi} \\
        \gamma_n &=& -2\sqrt{\frac{5}{6}}\beta\sqrt{S_n m^*} \geq 0 \\
        k_t &=& 8G^*\sqrt{R^*\xi} \\
        \gamma_t &=& -2\sqrt{\frac{5}{6}}\beta\sqrt{S_t m^*} \geq 0 \\
        S_n &=& 2Y^*\sqrt{R^*\xi}, S_t=8G^*\sqrt{R^*\xi} \\
        \beta &=& \frac{ln(e)}{\sqrt{ln^2(e)+\pi^2}} \\
\end{eqnarray}

The mixed variables are defined as:

\begin{eqnarray}
        \frac{1}{Y^*} &=& \frac{1-\nu_1^2}{Y_1}+\frac{1-\nu_2^2}{Y_2} \\
        \frac{1}{G^*} &=& \frac{2(2-\nu_1)(1+\nu_1)}{Y_1}+\frac{2(2-\nu_2)(1+\nu_2)}{Y_2} \\
        \frac{1}{R^*} &=& \frac{1}{R_1}+\frac{1}{R_2}, \frac{1}{m^*}=\frac{1}{m_1}+\frac{1}{m_2}
\end{eqnarray}

For this study the particles are identical with their parameters given by $Y=5\cdot10^6 Pa$,
$\nu=0.45$, $e=0.3$, and $R=d/2=0.5mm$. The mass is fixed by setting the
density equal to $2500 kg/m^3$. The friction coefficient is given three different
values $\mu=0.25, 0.5, 0.75$ in order to carry out a parametric study of the results
as a function of the properties of the medium. The friction coefficient is
especially influential because it determines the degree of re-arrangement between
contacting particles. As a consequence, beds of grains with a low value of $\mu$
become more compact and denser. Other parameters do not have this influence on the
structure of the bed. Thus, the role of the Young's modulus is to prevent grains
from interpenetrating each other. It may have a minor influence in the density
or packing fraction of the bed, but not in its microstructure. The coefficient of
restitution controls the time required to reach a static situation, but it cannot
prevent the rearrangements.

The parameter values selected give rise to results that are consistent other computational 
and experimental studies in the literature (see, \textit{e.g.},~\cite{Lix,Zhou,Yang,Wan,Darias,Gonzalez1,Li,Tian,Rubio,Rapaport,Gonzalez2}).
The value chosen for the Young's modulus ensures that the particles are stiff 
enough to display a realistic granular behaviour, while keeping the integration 
time step $\delta t$ at manageable values. In the present case $\delta t=10^{-5} s$ 
which is $20$ times smaller than the average duration of the contact of particles 
in the vicinity of the outlet size, as estimated from their velocities~\cite{Schafer}. 
However, some simulations have been repeated with an integration time step 
$\delta t=5\cdot10^{-6} s$ (and $\mu=0.5$)to ensure that the results hold.

Finally, we note that the particle-wall interaction is modelled in the same way as
particle-particle interactions, but assuming that the second particle has infinite
mass and radius (flat wall limit). The material properties of the second particle
are the same than those of the bulk particles.

\subsection{Simulation protocol}
\label{simulations}
The protocol used to obtain the results presented in this work is as follows. First,
particles are randomly placed inside the silo and given random velocities. Initially,
the outlet at the bottom is closed, and particles are allowed to settle under the
influence of gravity. The creation and settlement of particles is done in several batches 
until reaching the desired number of grains. Once the kinetic energy has achieved a 
negligible value, $E_k\le 10^{-10}J$, so particles can be considered to be at rest, 
the outlet is opened and particles outflow. In order to keep the conditions of the 
discharge constant, after the exiting particles have fallen a distance $4d$ they are
reintroduced at the top of the silo.

Although the silo discharge reaches a stationary state rather fast as monitored,
\textit{e.g.}, by the kinetic energy, the simulation is allowed to proceed during
several hundred thousands of time steps before recording data for analysis. To study
the collision frequency, $20000$ consecutive time-frames are used. This allows to
record several thousands of collision events. To study average quantities such as
the packing fraction and the typical velocities, around $1000$ time-frames separated
from each other by $10000$ time steps are used. When using the smaller integration
time step the frequencies are doubled.

For outlet sizes such that stable arches appear ($D\leq 5d$) we use the following
procedure to identify a clog and remove the arch. Based on trial simulations, when
the kinetic energy is below $10^{-10}J$ (for comparison, this is around four orders 
of magnitude lower than the steady state kinetic energy) it is safe to say that the 
flow is arrested. The particles forming the blocking arch are identified by locating 
those with lowest vertical coordinate whose horizontal position lies within the limits 
of the outlet. This particles are displaced vertically to a position below the base 
of the silo in one time step. This has always been observed to be enough to resume the
flow.

\begin{figure}
\begin{center}
\includegraphics[width=\textwidth, keepaspectratio=true]{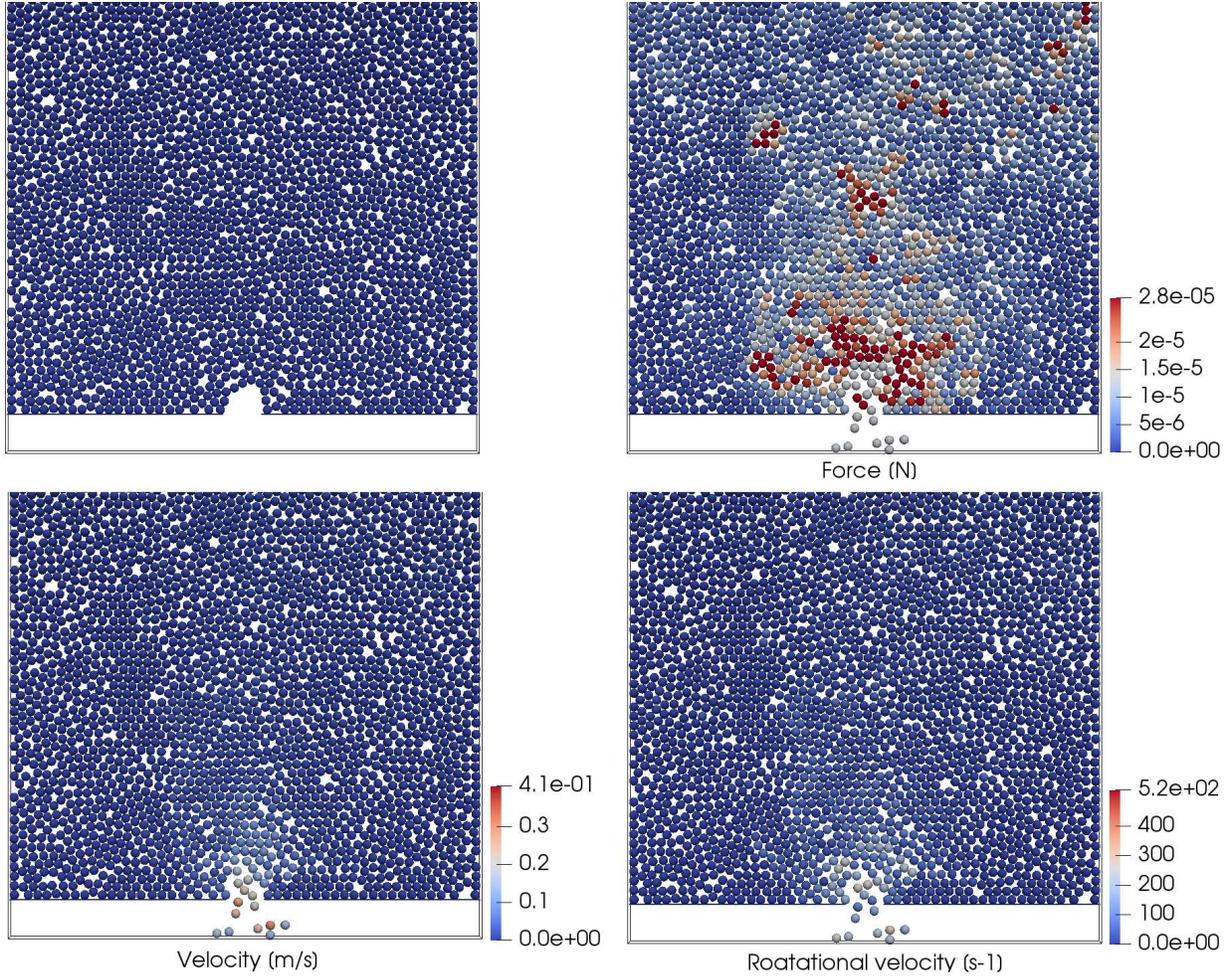}
\end{center}
\caption{{\bf Snapshots of a simulation with $D=4d$ and $\mu=0.5$.}
\textit{Top-Left}: An instance of an arch arresting the flow.
\textit{Top-Right}: Total force on each particle.
\textit{Bottom-Left}: Velocity magnitude.
\textit{Bottom-Right}: Rotational velocities.
}
\label{fig2}
\end{figure}

In Fig~\ref{fig2} are shown some snapshots of a simulation with $D=4d$ and
$\mu=0.5$. In the top-left panel there is an instance of an arch. Two particles
can be seen ``hanging'' due to friction. The other three panels show the same time
instant, taken a few time steps after the removal of the arch. The top-right panel
shows the total force on each particle. The forces distribution is very
heterogeneous, as is characteristic of granular media ~\cite{Behringer}. There
is a concentration of large forces on particles close to the exit, probably due
to particles precipitating to close the space left by the removal of the arch.
The bottom-left panel shows the velocity magnitude, while the bottom-left
panel shows the angular velocities. The linear velocities close to the outlet are
roughly below $0.2m/s$, decreasing fast in the bulk. The distribution of rotational
velocities is analogous to that of the linear ones.

\begin{figure}
\begin{center}
\includegraphics[width=\textwidth, keepaspectratio=true, angle=180]{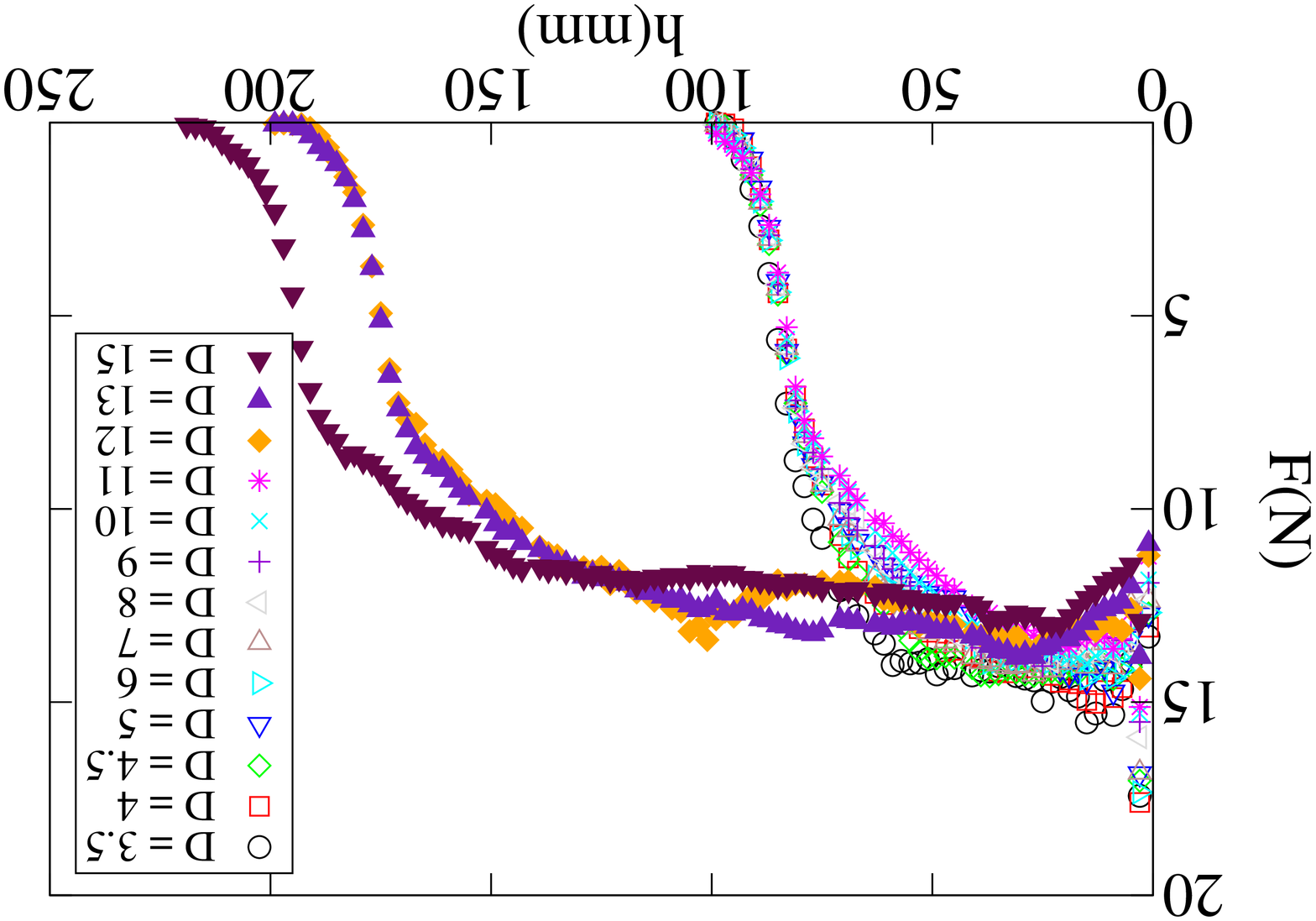}
\end{center}
\caption{{\bf Janssen Effect.}
Horizontal component of the force exerted on the vertical walls of the silo for
all outlet sizes. The data are averages over time during the discharge of the silo.
}
\label{fig3}
\end{figure}

The Janssen effect in our simulations is shown in Fig~\ref{fig3}. In particular, the plot
displays the horizontal force exerted on the vertical walls as a function of the height
during the discharge of the silo. The force is zero in the free surface of the silo but rises
fast as the height decreases. After a few tens of particle diameters the force reaches a
quasi-plateau in which the force increases very gently. This behavior is entirely similar to
that observed in other granular dynamical systems (see, \emph{e.g.}, ~\cite{Bertho,Windows}).
Note that all the signals tend to the same value of the force near the bottom, independently 
of the filling height. 
The peak observed near the bottom of the silo, that causes a dispersion of the data, is due 
to the sequential initial filling used. This effect is thoroughly discussed in~\cite{Ciamarra}.
In the present study particles are allowed to flow until a steady state develops, which erases 
the memory of the filling protocol.

\section{Results and Discussion}
\label{results}
\subsection{Collision rate}
\label{results_cr}

\begin{figure}
\begin{center}
\includegraphics[width=\textwidth, keepaspectratio=true]{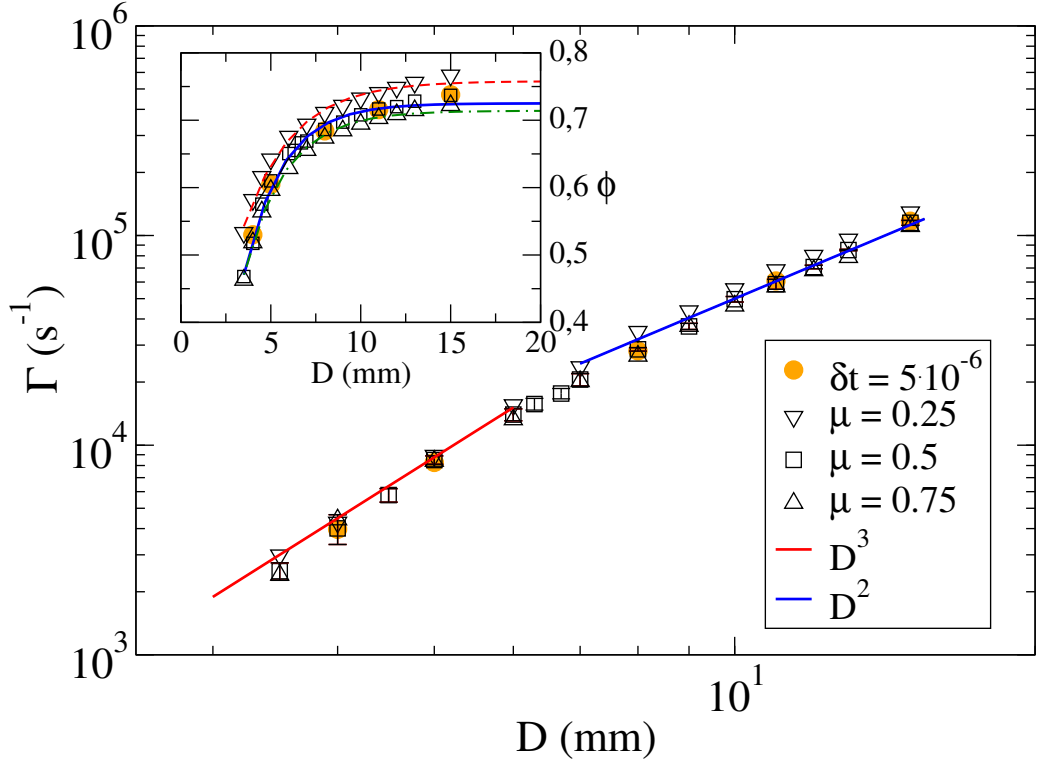}
\end{center}
\caption{{\bf Collision frequency.}
Collision frequency as a function of the outlet size. The open symbols
represent the main result, while the orange circles are the checking values
obtained reducing the integration time step of the simulations, and $\mu=0.5$. 
The red and blue lines are fitting lines. \textit{Inset}: packing fraction in 
the observation window as a function of the outlet size. The lines are a fit 
to function Eq.~\ref{packfract}}
\label{fig4}
\end{figure}

Since we are interested in the dynamics of particles close to the outlet, we focus
on a square area of size $D\times D$ whose bottom side coincides with the outlet of the
silo. More important, inside this window of observation is where most blocking arches
form, as has been demonstrated experimentally~\cite{Garcimartin2}. To count the
number of collisions, the contact matrix (a $N\times N$ matrix
whose values are $c_{ij}=c_{ji}=1$ if particles $i$ and $j$ are in contact and zero
otherwise) for all the particles
in the simulation is built at each time step. A collision is counted whenever two
particles are in contact in time step $t_i$ \textit{and} they were not in contact
in the previous time step $t_{i-1}$. When clogging appears all the collisions
occurred between the removal of one clog and the next are included in the calculation
of the collision frequency. Given that the area of the observation window
grows with $D$ one would anticipate that the collision frequency $\Gamma$ scales
as $D^2$. However, as can be seen in Fig~\ref{fig4} this is only the case for
outlet sizes $D\gtrsim 7d$. Below this value the collision frequency grows faster
than expected, with an exponent approximately equal to $3$. As in the case of the
mass flow rate, this can be understood taking into account that the packing fraction
$\phi$ of the particles in the area also grows with $D$. The inset to Fig~\ref{fig4} 
shows the packing fraction computed in the observation window. As has been observed in 
experimental systems~\cite{Zuriguel_obs,Janda} the growth behaves as a saturating 
exponential. The saturation is reached for a value slightly above $0.7$. The dependence 
can be well fitted to a function of the form~\cite{Mankoc,Janda}:

\begin{equation}
\phi\left(D\right)=a\left(1+be^{-D/c}\right).
\label{packfract}
\end{equation}

The fitting parameters are summarized in Table~\ref{parset}. The parameter $a$
is the asymptotic value of the packing fraction for large aperture sizes. As
discussed above, it decreases monotonically upon increasing the coefficient of
friction because higher values of $\mu$ prevent the rearrangement of particles in
the bed. The parameter $c$ has the dimensions of a distance and its value is
between $2d$ and $3d$, with no monotonous dependence on $\mu$. This is consistent
with the experimental value of ``diffusive constants'' derived from fitting the vertical
velocity profile measured inside the silo ~\cite{Choi1,Choi2,Zuriguel_fluct, Garcimartin}
to the phenomenological continuous model derived in~\cite{Tuzun}.

\begin{table}
\centering
\caption{Parameters from fitting the data in Fig~\ref{fig4} and
        Fig~\ref{fig5}a,b,c to Eq.~\ref{packfract}}
\label{parset}
\begin{tabular*}{\columnwidth}{@{\extracolsep{\fill}}cccc@{}}
\hline
$\mu$ &  0.25 & 0.5 & 0.75 \\
\hline
\multicolumn{4}{c}{Parameters from packing fraction Fig~\ref{fig4}} \\
\hline
$a [-]$             & 0.758  & 0.725  & 0.714   \\
$c [mm]$            & 2.74   & 2.14   & 2.32    \\
\hline
\multicolumn{4}{c}{Parameters from collision frequency Fig~\ref{fig5}a} \\
\hline
$a [s^{-1}mm^{-2}]$             & 759.8  & 698.7  & 670.5   \\
$c [mm]$            & 1.85   & 2.65   & 2.15    \\
\hline
\multicolumn{4}{c}{Parameters from collision frequency Fig~\ref{fig5}b} \\
\hline
$a [s^{-1}mm^{-2}]$             & 783.2  & 726.9  & 697.3   \\
$c [m/s]$           & 0.073  & 0.075  & 0.053   \\
\hline
\multicolumn{4}{c}{Parameters from exit velocity Fig~\ref{fig5}c} \\
\hline
$\gamma [-]$        & 1.097  & 0.993  & 0.871   \\
\hline
\end{tabular*}
\end{table}

\begin{figure}
\begin{center}
\includegraphics[width=0.5\textwidth, keepaspectratio=true]{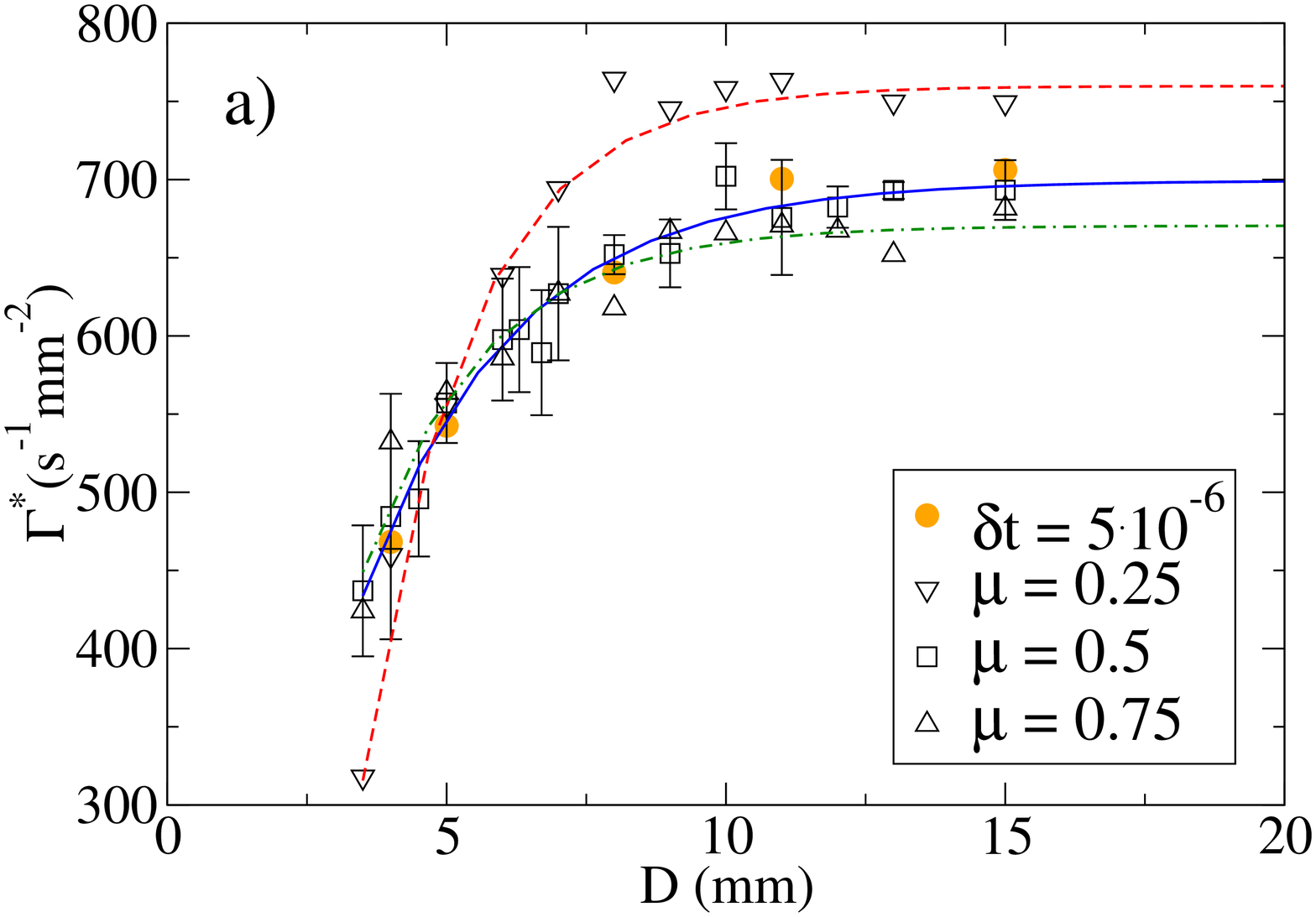}
\includegraphics[width=0.5\textwidth, keepaspectratio=true]{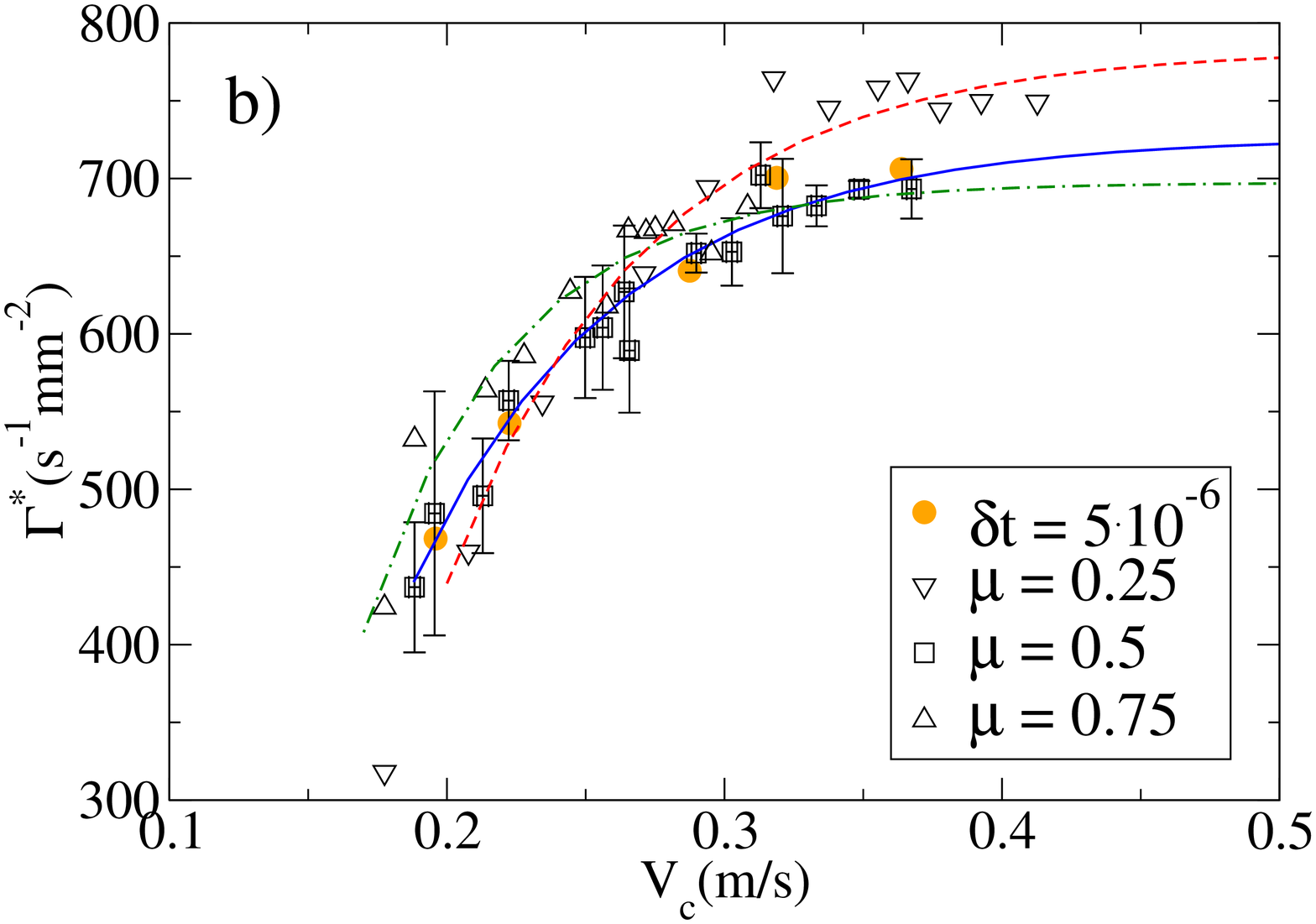}
\includegraphics[width=0.5\textwidth, keepaspectratio=true]{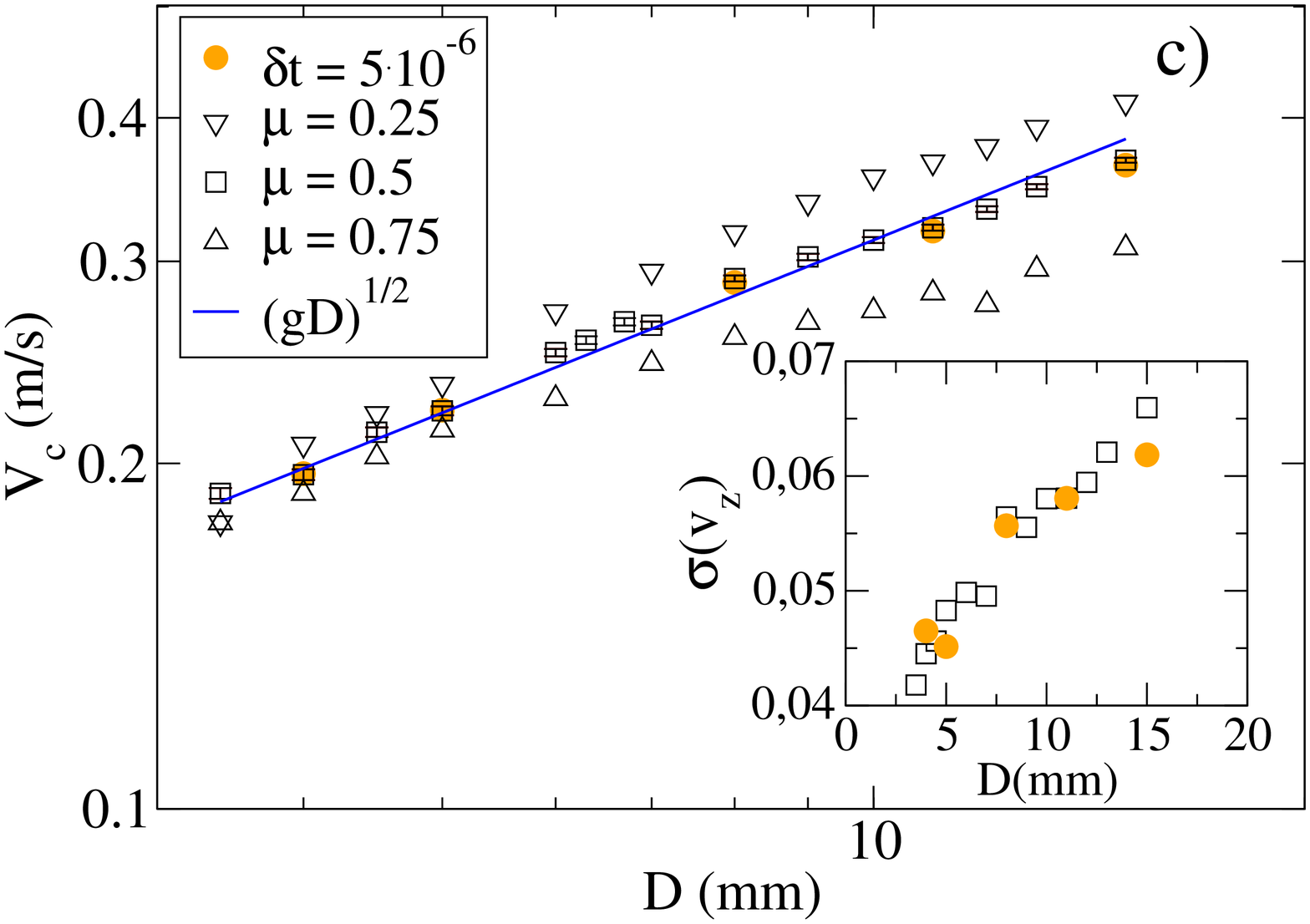}
\end{center}
\caption{{\bf Normalised collision frequency.}
\textit{Top}: Normalised collision frequency as a function of the outlet size. The
empty symbols represent the main result, while the orange circles are the checking
values obtained reducing the integration time step of the simulations and $\mu=0.5$.
\textit{Middle}: Normalised collision frequency as a function of the vertical
velocity of the particles exiting through the center of the outlet. The continuous
lines are fits to function Eq.~\ref{packfract}. \textit{Bottom}: Vertical velocity
of the particles exiting through the center of the outlet as a function of the
outlet size. \textit{Inset}: Variance of the velocity of particles inside the
observation window.
} \label{fig5}
\end{figure}

A normalised collision frequency is defined dividing the frequency by the area of
the window of observation and the packing fraction $\Gamma^*=\frac{\Gamma}{D^2\phi}$. In
this way we take into account the increasing size of the window of observation and the
varying packing fraction. Note that $D^2\phi$ is approximately the number of particles
inside that window. This normalised collision frequency is plotted in
Fig~\ref{fig5}a. The results obtained with the three values of $\mu$ are reported
but only the error bars of the case $\mu=0.5$ are shown, the others being
similar. The collision frequency is found to grow for small
values of the outlet size. The rate of growth starts to decrease around $D=7d$ and
it finally reaches a saturating value near $D=10d$. The data can be fitted to a
function of the same form than Eq~\ref{packfract}. In
this case the value of the characteristic length is $c\approx 2d$ again (see
Table~\ref{parset}), which is
consistent with the value derived from the behaviour of the packing fraction. Note
that the lower the packing fraction of the bed the lower the value of the normalised
collision frequency, since there are fewer particles to provoke collisions. This
trend is observed respect to the friction coefficient as well: for a high value
of $\mu$ the curve is lower because the bed is less dense, while for the lower
value of $\mu$ the curve moves upwards. It should be noted that the dependence of
the collision frequency on the outlet size is the same if one uses
$\Gamma^*=\Gamma/D^2$ instead of $\Gamma^*=\Gamma/\left(\phi D^2\right)$, only the
values change. So, this dependence is not an artefact of dividing by $\phi$.
The formal similarity of the curves in Fig~\ref{fig4} and Fig~\ref{fig5}a strongly
suggests a relationship of the former result to the behaviour of the packing
fraction, although other effects might play a role.

In Fig~\ref{fig5}b the normalised frequency is plotted against the vertical
velocity of the particles outflowing through the centre of the outlet
$V_c=V_z(W/2,0,0)$. This velocity is chosen as a characteristic velocity because
the vertical velocity of the particles changes inside the $D\times D$ window as they
fall. The advantage of $V_c$ is that it has a predictable value~\cite{Rubio,Janda}
very approximately given by $\sqrt{\gamma gD}$ where $g$ is the acceleration of
gravity and $\gamma$ is a parameter whose value is close to $1$. As can be seen
in Fig.~\ref{fig5}c the fit is very close for $\mu=0.5$, while the exit velocity is
slightly higher and slightly lower for lower and higher friction, respectively.
Additionally, the range of values spanned by $V_c$ is wider than that of the
average vertical velocity in the box or the average speed, which makes the result
easier to visualize. Although the saturation is not as clear as in the case of
$\Gamma^* vs D$, the data-points are fitted with the same saturating exponential
Eq~\ref{packfract}. The characteristic velocity given by the fitting parameter
$c$ is ${V_c}^*\approx 0.075 m/s$. This value corresponds with those experimentally 
found in the bulk of the silo~\cite{Fullard,Zuriguel_fluct,Garcimartin,Gonzalez2,Benyamine}.
Additionally, the values of $\gamma$ further stress consistency with experiments~\cite{Rubio,Janda}.

This result is very surprising. Intuitively, one would expect to see an increase
in the collision rate of the particles upon increasing their velocities
~\cite{Falcon,Aumaitre}. Although the silo is an open system we have seen that
the space available for particles shrinks upon increasing $D$. However, collision
rate increases only up to moderate values of the outlet
size $D\lesssim 7d$. Onwards, the collision rate flattens in spite of the velocity's
continuous increase. Interestingly, the so called ``granular temperature'', given
by the variance of the velocity~\cite{Lun}, also increases monotonically (see inset
to Fig~\ref{fig5}c) in a linear fashion. This means that the agitation of the
particles is not enhancing the probability of collisions either. It appears that
the collision rate only increases while the area around the outlet is being filled
with particles, as measured by the packing fraction in order to compare the increasing areas.

The new result shown in Fig~\ref{fig5}a,b has implications for the probability
of arch formation and clogging. The number of particles in a clogging arch grows
linearly with $D$~\cite{Garcimartin2}.The chance encounter of a growing number of
particles should hence decrease with $D$. This intuition is made a strong argument
in~\cite{Durian_prl} where it is
shown that the number of configurations of grains near the exit that cause a clog
falls exponentially upon increasing the outlet size. Hence, the flow needs to
sample an exponentially large number of grains' configurations before finding one
able to block the exit. Our new result implies that this sampling is hindered by
the saturation in the collision rate that results from filling with particles the
area near the outlet. This strengthens the conclusion than when increasing the
outlet size clogs are possible but extremely unlikely.

Furthermore, the continuity of the curve $\Gamma^* vs D$ suggests that there is
no critical outlet size. Instead, the transition from the clogging to the continuous
flow regime is smooth.

\subsection{Distribution of velocities}
\label{reuslts_sv}

\begin{figure}
\begin{center}
\includegraphics[width=0.5\textwidth, keepaspectratio=true]{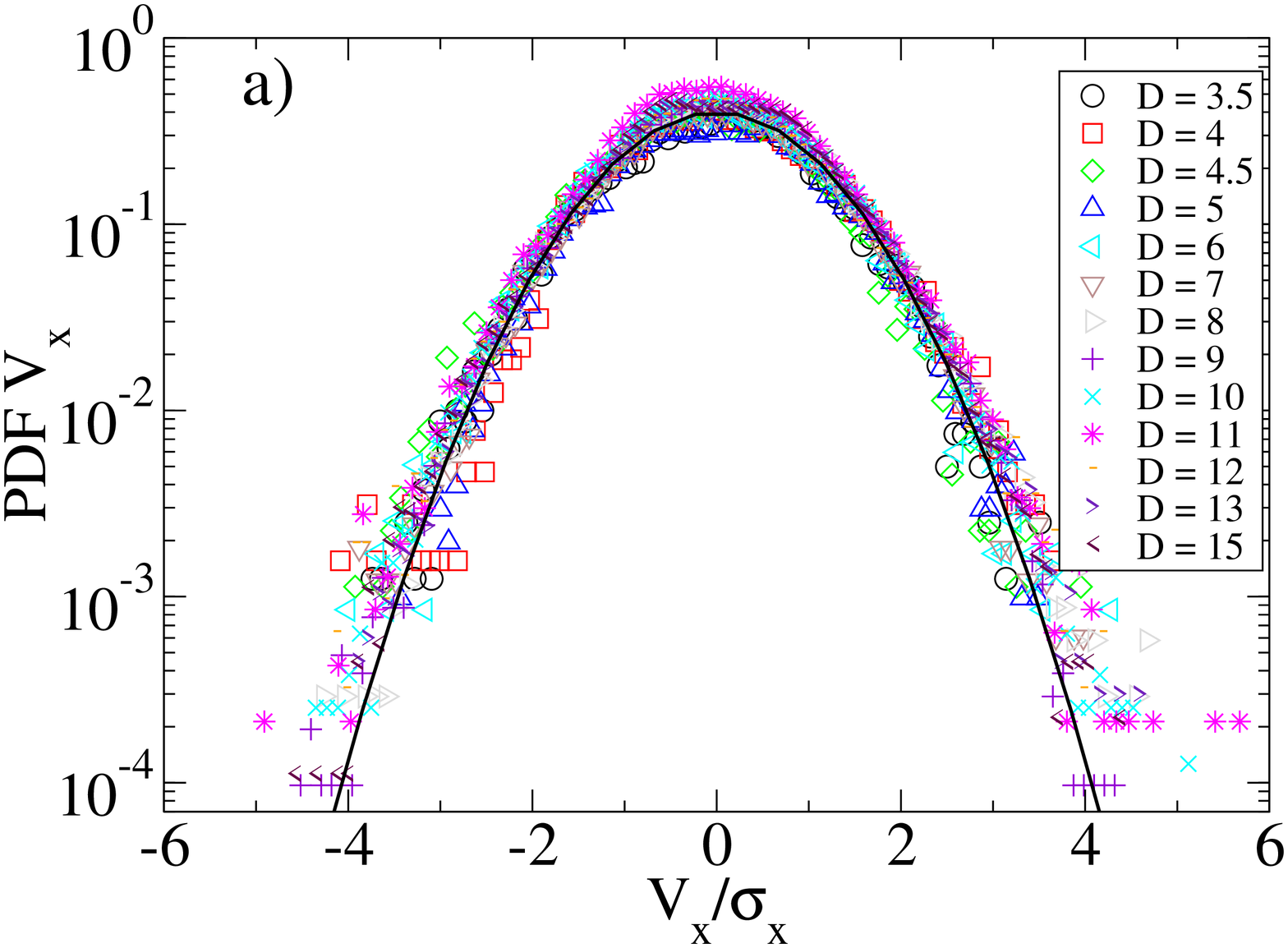}
\includegraphics[width=0.5\textwidth, keepaspectratio=true]{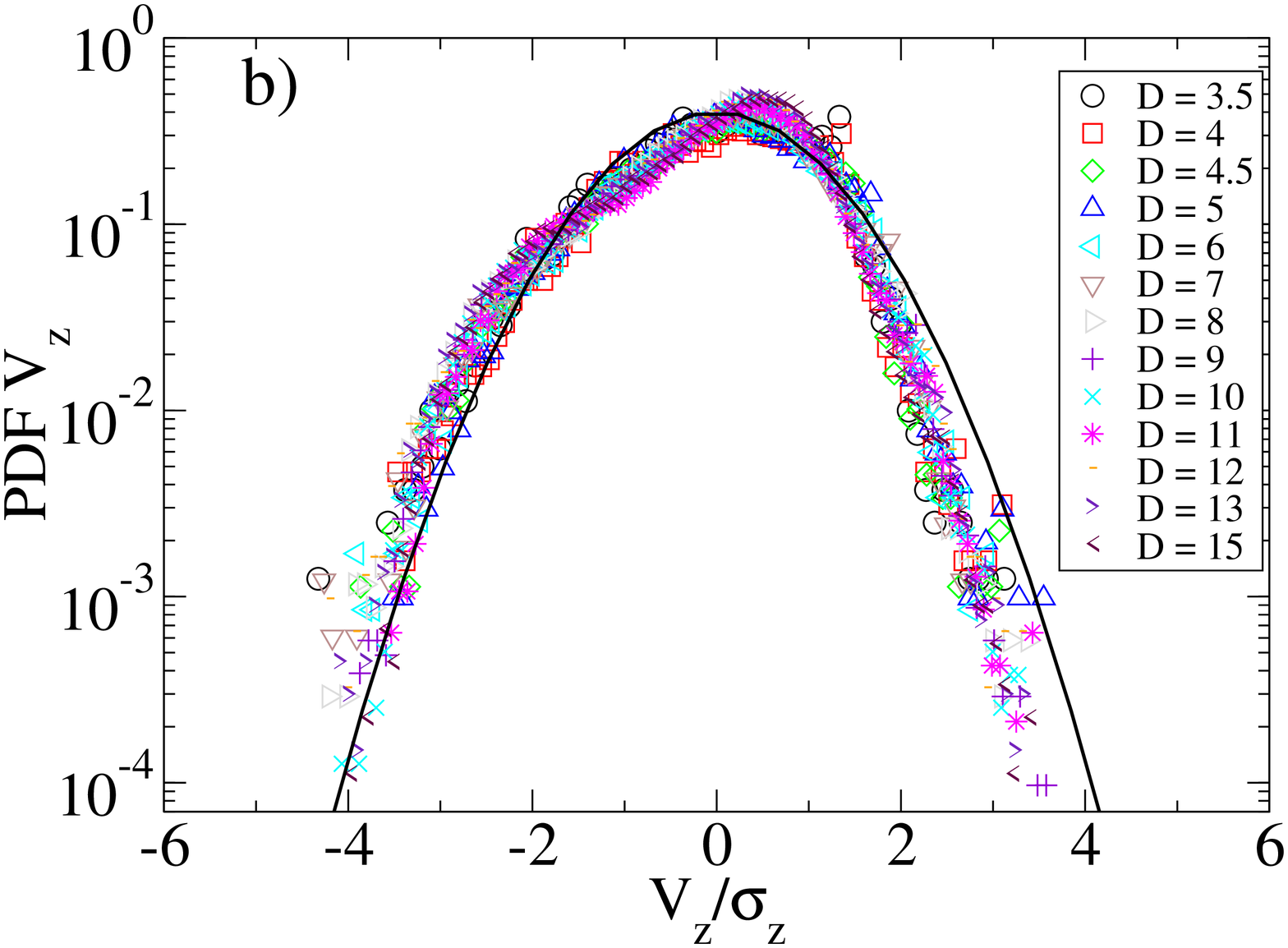}
\includegraphics[width=0.5\textwidth, keepaspectratio=true]{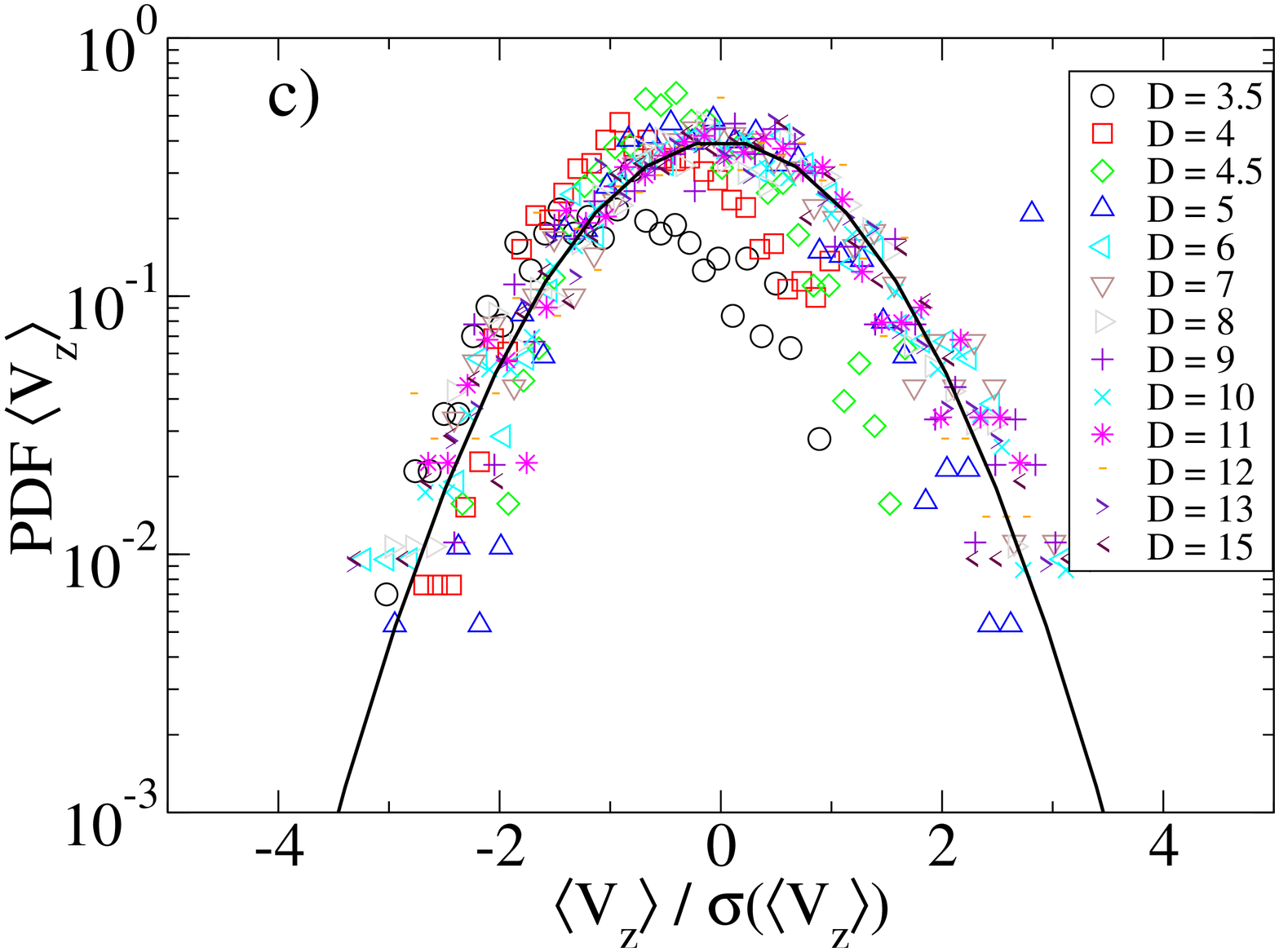}
\end{center}
\caption{{\bf Velocity histograms.}
\textit{Top}: Probability distribution function of the horizontal velocities
of individual particles. \textit{Middle}: Probability distribution function of the
vertical velocities of individual particles, subtracting the average in the window.
\textit{Bottom}: Probability distribution function of the fluctuations of the average
vertical velocity computed in the window of observation. The distributions are
normalised and the velocities rescaled by the standard deviation. The black line is
Gaussian function with average equal to zero and standard deviation equal to the unity.
There is no fit in the plots.
} \label{fig6}
\end{figure}

In order to explore further the dynamics of the colliding particles near the exit,
we compute the probability distribution functions (PDF) of the horizontal and vertical
instantaneous velocities. For clarity of the plots we report values for $\mu=0.5$.
These are shown in Fig~\ref{fig6}a and Fig~\ref{fig6}b, respectively with the velocities 
normalised by their standard deviation. The distribution of horizontal velocities 
follows excellently a Gaussian shape for all values of $D$. Note that
there is no fit to the data due to the normalization applied. This result is natural
since there are no forces or other influences in the horizontal direction. To build
the PDF of the vertical velocities, the average is computed inside the window of
observation and subtracted from the instantaneous velocity of each particle. This
PDF deviates from a Gaussian in two ways. First, the tails of the distribution show
an overpopulation of downwards (negative) velocities and a depletion of upwards
(positive) velocities. This result is natural given that most particles are
effectively falling and that the average is not constant inside the window (it
varies with the vertical coordinate). Second, close to the centre of the
distribution the opposite trend appears: the positive velocities are enhanced while
the negative ones are depleted. This is due to the collisions between the
particles. Effectively, when particles enter the observation window their
accelerations and velocities are minimum~\cite{Rubio}, hence collisions are more
likely to produce large deviations in the velocity vectors. Whereas when particles
have fallen some distance inside the window and built up greater acceleration, they
are more difficult to deviate.

Previous works~\cite{Moka,Arevalo,Garcimartin} have reported non-Gaussian velocities
for individual particles in a silo. All those works carried out their measurements
in the bulk of the silo, instead of in the vicinity of the exit. As has been
discussed in the introduction the dynamics in that region is dominated by lasting
contacts due to the larger density of the medium. This enhances the correlations
between particles~\cite{Garcimartin} which give rise to deviations from the
Gaussian in the PDFs of velocities and displacements. Indeed, in~\cite{Garcimartin} 
it can be seen that as one approaches the bottom of the silo the non-Gaussian features 
of the distribution of particles' velocities wear off. Comparing Fig~\ref{fig6} with the 
experimental results in, \emph{i.e.},~\cite{Choi2,Zuriguel_fluct,Garcimartin} one 
can see that, again, the presented simulations are consistent with the experimentally 
measured behavior of particles in a silo. 

In Fig~\ref{fig6}a,b the velocities refer to individual particles. We also compute
the average over particles in the $D\times D$ window during several seconds. This average
fluctuates over time but has a well-defined stationary value. Subtracting the
stationary value from the instantaneous average in each time step we obtain the
fluctuations of the average velocity. This is displayed in Fig~\ref{fig6}c for the
vertical component. The distribution is Gaussian in all cases, only showing
deviations for the smallest values of outlet $D\leq 4.5d$. Note that for these values the 
clogs are so frequent that the time-series to be used to produce the histograms are 
very short. Besides, the data present a rather wide oscillation inside each time-series. 
Both effects, preclude obtaining a well-defined average. However, the result shown is 
consistent with the measurements presented in~\cite{Garcimartin}, where the fluctuations 
of the average velocity magnitude are found to be Gaussian. The deviations in the result
for small orifices is due to lack of statistics. Effectively, the size of the window
is very small in these cases, housing a very small number of particles compared
to larger apertures.

The conclusion presented at the end of section~\ref{results_cr} regarding the
absence of evidence of a critical outlet size is further reinforced by the results
just discussed. The PDFs shown in Fig~\ref{fig6}, once normalised, are all the same
for small and large apertures. In the horizontal direction they are perfectly
Gaussian, as expected. The deviation from gaussianity in the vertical component of
individual particles can be explained by the collisional dynamics in the vicinity
of the outlet. Finally, the fluctuations of the average are Gaussian, in agreement
with experiments, for all outlets. This indicates that the dynamics of particles
in the region near the exit, where blocking arches form, is unaffected by the
increasing area of the outlet.

\subsection{Distribution of time lags}
\label{lagtimes}

\begin{figure}
\begin{center}
\includegraphics[width=0.5\textwidth, keepaspectratio=true]{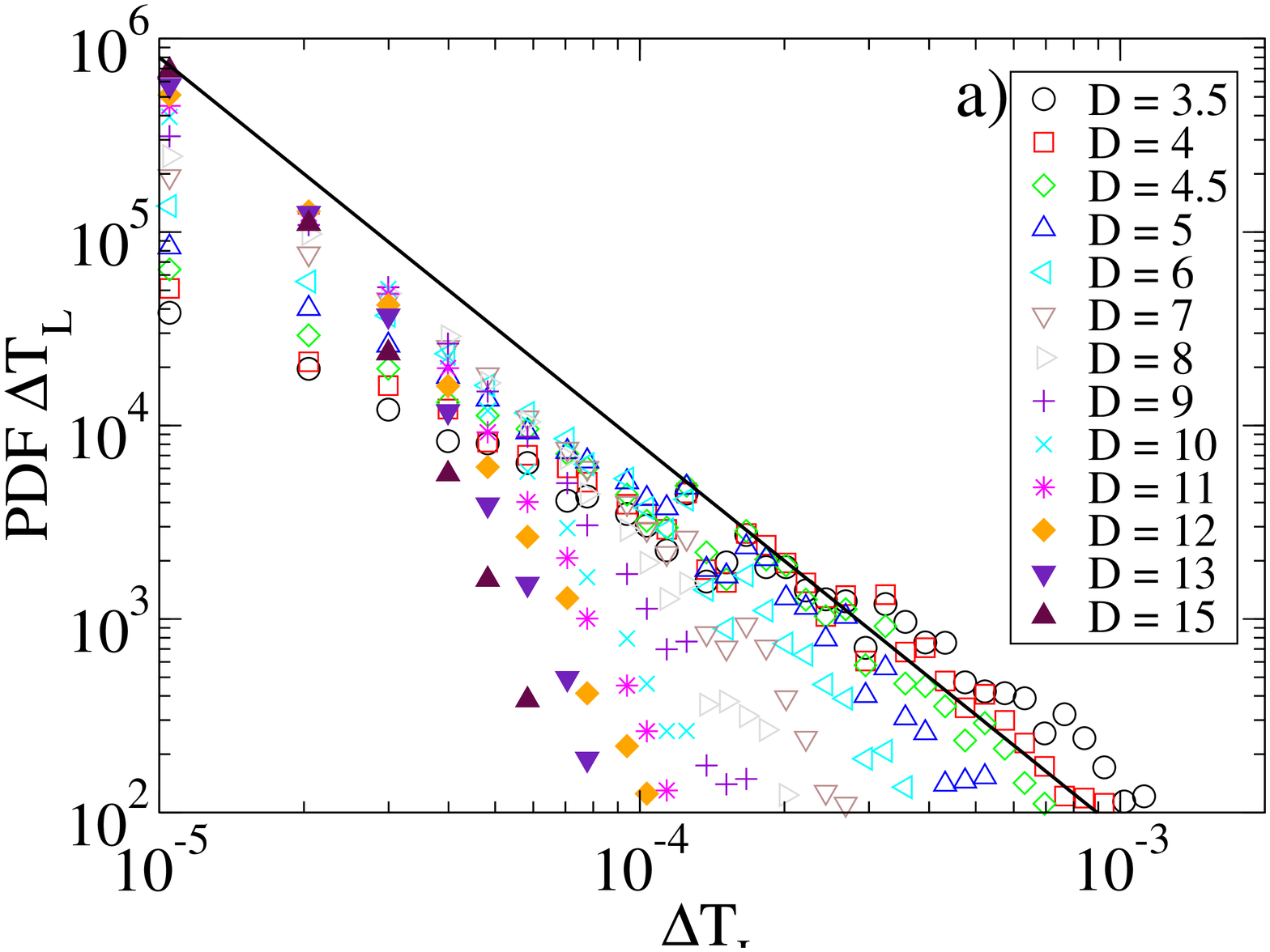}
\includegraphics[width=0.5\textwidth, keepaspectratio=true]{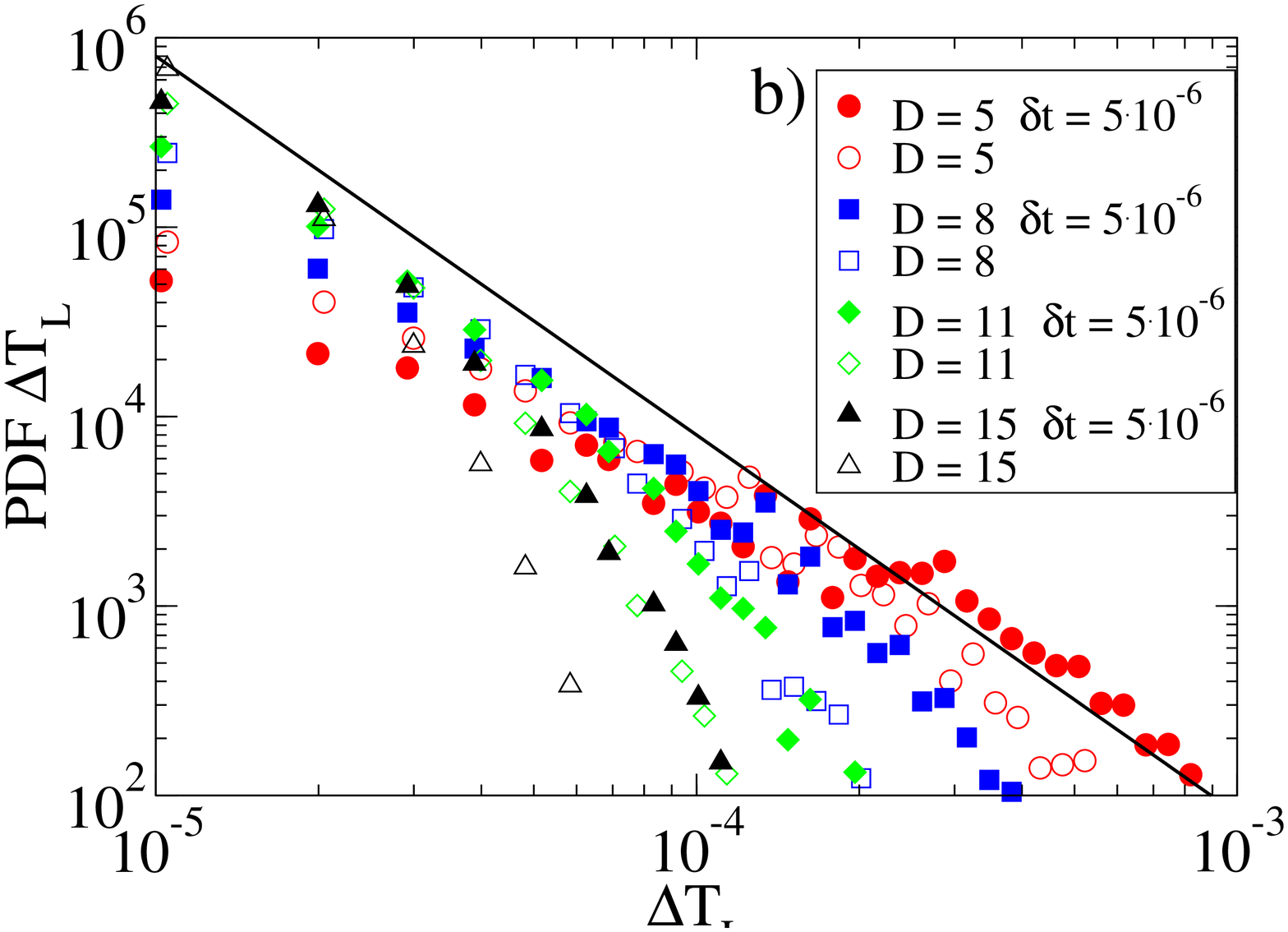}
\includegraphics[width=0.5\textwidth, keepaspectratio=true]{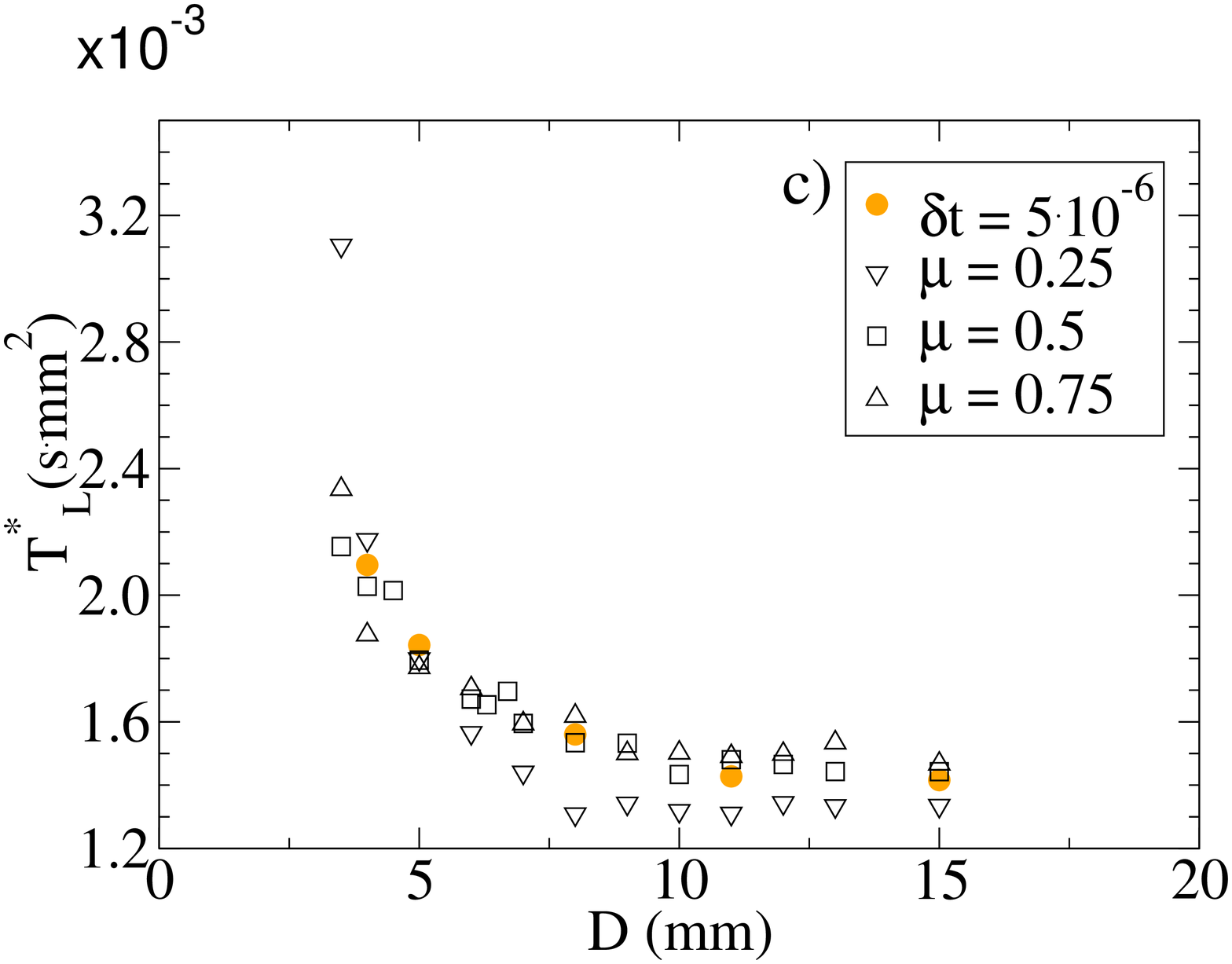}
\end{center}
\caption{{\bf Time lag statistics.}
\textit{Top}: Probability distribution function of the time lags between
successive collisions. \textit{Middle}: Same distributions obtained from
simulations using a shorter integration time step and compared with the
originals displayed in the left panel. The black lines are guides to the eye with a
slope equal to $-2$. \textit{Bottom}: Average normalised time lag
$\Delta T_{L}^*=\Delta T_L\times D^2\times \phi$ as a function of the outlet size.
} \label{fig7}
\end{figure}

We now study the time lag $\Delta T_L$ between successive collisions. The PDFs for
all cases with $\mu=0.5$ are shown in Fig~\ref{fig7}a. The distributions roughly
follow a power law distribution falling with an exponent equal to $2$. However,
there are differences.
First, short time lags are more likely in the cases of large apertures. Second,
the trend reverses for larger $\Delta T_L$, which are less likely for large $D$.
This observation has an intuitive explanation based on the results presented for
the packing fraction (see Fig~\ref{fig4}). In the case of a small window of
observation the packing fraction is low and particles can travel longer distances
between collisions. As $D$ increases, so does the packing fraction and,
consequently, the time between successive collisions shortens. Additionally, the
velocity of the particles is lower for lower values of $D$ (see Fig~\ref{fig5}c).

In Fig~\ref{fig7}b we show the time lag distributions for a few values of $D$ for
which simulations were carried out with a reduced integration time step. The values
displayed change slightly, but the general conclusions remain unchanged. Reducing
the integration time increases the probability to observe large time lags. Instead,
the frequency of short lag times is corrected by a very small amount.

The behaviour of the normalised average time lag $T_{L}^*=T_L\cdot D^2\cdot \phi$ as
a function of the aperture size and the friction coefficient is shown in
Fig~\ref{fig7}c. As deduced from the histograms, the time lag is larger for small
values of $D$ and decreases upon increasing $D$. More interesting, there is a
saturation for $D\gtrsim 10d$ above which the curve flattens. This behaviour
is concomitant with the saturation of the packing fraction and the collision
frequency discussed earlier. The influence of the friction coefficient on the
curves is reversed with respect to the collision frequency curves.

\section{Conclusion}

This work analyses the behaviour of particles in a silo in the region close to the
outlet. The dynamics in this region is characterised by a collisional regime, in
which contacts are short-lived. This is in contrast with the bulk of the silo where
contacts are permanent or long-lived. Interestingly, it is in this region where
the arches that block the exit and arrest the flow (for outlet sizes $D\lesssim 5d$)
are formed \cite{Garcimartin2}. The study covers a range of outlet sizes $D$ from
small ones, where arches
appear frequently and is necessary to remove them, to big apertures for which
flow is continuous. It has been shown that the collision frequency of the particles
increases with $D$ until it reaches a saturation value and flattens from
$D\approx 10d$ onwards. This is coincident with the saturation of the packing
fraction which, as had been previously observed, behaves similarly. The same
behaviour is obtained when the collision frequency is studied as a function of the
velocity of the particles. This is puzzling given that the velocity increases
continuously and one would expect to see more frequent collisions upon greater
velocities \cite{Falcon,Aumaitre}.

The main result holds when the friction coefficient of the particles changes. This
material coefficient results in a denser packing when it is small because it allows
particles to rearrange. When the friction takes high values particles get interlocked
more easily and the density of the bulk medium decreases. The consequence for the collision
frequency is that it increases for the lower value of $\mu$ because the chances of
encounter between particles increase. Instead, when the friction coefficient is
raised the collision frequency decreases because particles have more space available
to move.

The saturation of the collision frequency has consequences for the existence of a
critical outlet size from which arches never appear. In the frame of the statistical
argument put forward in \cite{Durian_prl} the flattening of the collision frequency
implies a reduced number of chance encounters between particles and, consequently,
a reduced ability of the flow to sample configurations compatible with a clog.

Furthermore, the analysis of the distribution functions of the velocities shows
no transition in respect to the outlet size. Tellingly, the histograms of the
individual velocities are gaussian for both horizontal and vertical components and
for all values of $D$. The small deviations present in the histogram of the vertical
velocity (the average subtracted) are easily accountable by the dynamics of
collisions. The PDF of the fluctuations around the stationary value of the
average (over particles) vertical velocity is a gaussian as well, consistently with
experiments \cite{Garcimartin2}.

To complete the study of the collisional regime, the time lag has been analysed.
It is found that the time lag distribution follows roughly a power law decay with
exponent $2$. This general trend is nuanced by a depletion of short time lags
for small apertures and, conversely, a depletion of long time lags for large
outlet sizes. This is consistent with the behaviour of the packing fraction and the
collision frequency. Interestingly, the average time lag does not just decrease,
upon increasing $D$, it reaches a plateau in a similar (but inverted) fashion as the
collision frequency.

In conclusion, the collisional dynamics of particles does not show evidence of
critical behaviour respect to the outlet size. Furthermore, it is worth noting that
the change of behaviour (the transition to a plateau) of the collision frequency
and the time lag occur at a value of $D\gtrsim 7d$ larger
than that for which arches start to appear $D\lesssim 5d$. It is difficult to
harmonise this observation with the existence of a critical $D$ which separates
the continuous flow regime from the clogging regime. It is consistent, instead,
with a gradual decrease in the probability to observe clogs as the size of the
outlet increases.

The next logical extension of the present work would be to run simulations
in $3D$ to study the influence of dimensionality. Many of the characteristics
of the flow of granular matter inside a silo are analogous in $2D$ and $3D$,
including the existence of clogging and continuous regimes, the variation of
the flow with the outlet size, the variation of the packing fraction, the saturation
of pressure, the velocity profiles of the particles and the PDF of the velocities.
However, dimensionality could have an influence in the existence or not of some
behaviors. In this regard, it has been suggested~\cite{Janda_epl} that this variable
could have an influence on the existence of a putative critical outlet size.

The results in the literature show that both in $2D$ and $3D$ the outlet size defines
the relevant length scale and that it can be conveniently made dimensionless by the
typical size of the flowing particles. Having a dimensionless number allows to carry
the results to systems of different size. Effectively, the engineering literature shows
that interesting magnitudes such as the flow, pressure and velocities behave in the same
way than in laboratory setups or simulations. Hence, the results reported in this work
should be observable in industrial-scale silos because they depend on the $D/d$ ratio of
outlet to particle size. Of course, for practical reasons, industrial-scale silos usually
operate in the regime $D\gg d$, precisely to avoid clogging-related problems.

\begin{acknowledgments}

This work has been supported by \textit{EU} $H2020$ program project
\textit{BAMBOO} (GA 820771). I thank my colleague Ana Gonz\'alez-Espinosa
for critical reading of an early version of the manuscript.

\end{acknowledgments}

\bibliography{mybib.bib}

\end{document}